\documentclass[submission,copyright,creativecommons]{eptcs}
\providecommand{\event}{VPT 2017} 
\usepackage{breakurl}             
\usepackage{underscore}
\usepackage[utf8]{inputenc}
\usepackage{graphicx}
\usepackage{listings} 
\usepackage{tikz} 
\usepackage{caption,subcaption}   

\usepackage{multirow}
\usepackage{tabularx}
\usepackage{hhline}

\usetikzlibrary{arrows,shapes,positioning}

\lstdefinestyle{invis}
{numbers=none,frame=none,backgroundcolor=\color{white},mathescape,language=c}

\newcommand{\varname}[1]{{{#1}}}
\newcommand{\varvalue}[1]{{\footnotesize \textsf{#1}}}

\title{Towards Evaluating Size Reduction Techniques \\ for Software Model Checking}
\author{Gyula Sallai$^{1}$ \qquad \'Akos Hajdu$^{1,2}$\thanks{\thanksntp} \qquad Tam\'as T\'oth$^{1}$\thanks{\thanksrichter} \qquad Zolt\'{a}n Micskei$^{1}$
	\institute{$^1$Department of Measurement and Information Systems \\
		Budapest University of Technology and Economics, Budapest, Hungary}
	\institute{$^2$MTA-BME Lend\"ulet Cyber-Physical Systems Research Group, Budapest, Hungary}
	\email{salla@sch.bme.hu, \{hajdua,totht,micskeiz\}@mit.bme.hu}
}

\newcommand{\thanksrichter}{Partially supported by Gedeon Richter's Talentum Foundation (Gy\"{o}mr\H{o}i \'ut 19-21, 1103 Budapest, Hungary).}
\newcommand{\thanksntp}{Partially supported by Nemzeti Tehets\'eg Program, Nemzet Fiatal Tehets\'egei\'ert \"Oszt\"ond\'ij 2016 (NTP-NFT\"O-16).}

\def\titlerunning{Towards Evaluating Size Reduction Techniques for Software Model Checking}
\def\authorrunning{Gy. Sallai, \'{A}. Hajdu, T. T\'{o}th, Z. Micskei}
\begin{document}
\maketitle

\begin{abstract}
	Formal verification techniques are widely used for detecting design flaws in software systems.
	Formal verification can be done by transforming an already implemented source code to a formal model and attempting to prove certain properties of the model (e.g.\ that no erroneous state can occur during execution).
	Unfortunately, transformations from source code to a formal model often yield large and complex models, making the verification process inefficient and costly.
	In order to reduce the size of the resulting model, optimization transformations can be used. Such optimizations include common algorithms known from compiler design and different program slicing techniques.
	Our paper describes a framework for transforming C programs to a formal model, enhanced by various optimizations for size reduction.
	We evaluate and compare several optimization algorithms regarding their effect on the size of the model and the efficiency of the verification.
	Results show that different optimizations are more suitable for certain models, justifying the need for a framework that includes several algorithms.
	
	\vspace{1em}
\end{abstract}

\section{Introduction}
\label{sec:introduction}

As our reliance upon safety-critical computer systems grows, so does our natural desire for reliable proofs of their fault-free behavior. Such proofs can be given by formal verification algorithms, such as \emph{model checking}. Unlike testing, these are not only able to prove the presence of errors, but their absence as well, thus they are able to give a satisfactory answer on the safety of a system.

While there are several formal verification techniques available, incorporating them into a development workflow can pose a challenge. In model-driven development, we first design a model that describes our system, then we use formal verification to prove its safety and finally we create an implementation based on the defined (and safe) model. However, designing and defining a model for a project can be rather difficult and in many cases the financial and time constraints do not permit it. 

A possible solution to overcome this problem is to transform the already implemented source code to a formal model. However, a drawback of this approach is the large size of the model generated from the source code. As most verification algorithms have a rather demanding computational complexity (usually operating in exponential time and beyond), the resulting model may not admit efficient verification. A way to resolve this issue is to reduce the size of the generated model using \emph{optimizing transformations}.

This paper describes a framework which implements a \emph{transformation workflow} from C programs to a formal model, known as control flow automaton. This process is enhanced by some common optimization transformations, usually known from compilers. These simplify the model and reduce its size. Then the model is split into several smaller, more easily verifiable chunks using \emph{program slicing}~\cite{Weiser81}. As a result, the verification algorithm has to handle several smaller problems instead of a large one.

In this paper, we evaluate the effects of such transformations on the size of the model and the efficiency of verification. We use the compiler optimization algorithms known as \emph{constant propagation}, \emph{constant folding} and \emph{dead branch elimination}~\cite{DragonBook}. We also present and compare different program slicing techniques, namely \emph{backward slicing}~\cite{Ferrante87}, \emph{thin slicing}~\cite{Sridharan07} and \emph{value slicing}~\cite{Kumar15}. The models obtained after these transformation passes are then verified using \emph{predicate abstraction} in conjunction with the \emph{counterexample-guided abstraction refinement} (CEGAR) model checking algorithm. Our results show that the optimizations may allow significant reduction in the size of the model and the execution time of the verifier algorithm. However, the relative effectiveness of the algorithms varies between different input models, which justifies the need for a configurable framework that supports several optimization and slicing algorithms.

The rest of this section discusses related work. Section~\ref{sec:background} introduces the preliminaries of our work. Section~\ref{sec:workflow} presents our transformation workflow from C programs to formal verification along with some implementation details. Section~\ref{sec:evaluation} evaluates the transformations and their effect on verification. Finally, Section~\ref{sec:conclusions} concludes our work.

\paragraph{Related work.}
Predicate abstraction~\cite{graf97}, used in conjunction with CEGAR~\cite{clarke03} is a widely used technique for model checking software~\cite{SLAM,Blast,SATABS,SLAB,CPAChecker}.
However, the performance of these algorithms greatly depends on the size and complexity of the input model and source-to-model transformations tend to produce large models.
Several size reduction methods exist, the most prominent of them is program slicing~\cite{Weiser81, Ferrante87}.
Several other variants of slicing have been proposed since then, including thin~\cite{Sridharan07} and value slicing~\cite{Kumar15}. 
The software model checking tool SLAB~\cite{SLAB} successfully incorporates slicing into its workflow, allowing slicing to be performed on the abstract states.
In our tool we handle slicing and verification separately and use the former as a preprocessing step.
There has also been work on evaluating and comparing software model checkers \cite{Beyer16smt, Beyer15refinement}. However, these papers target the model checking algorithms, whereas the primary focus our work is on the size reduction techniques.
\section{Background}
\label{sec:background}

In this section we give a brief introduction to the theoretical background of formal verification, program representations, dependency structures and some compiler optimization algorithms used in our work.

\subsection{Formal Verification}
\label{sec:background:formalverification}
Model checking is a formal verification technique of mathematically proving correctness or faultiness of computer programs by systematically exploring their state space.
There are several program representations suitable for model checking. In this paper we use a formalism known as \textit{control flow automaton} (CFA)~\cite{Beyer09}. A CFA is a 4-tuple $(L, E, \ell_0, \ell_e)$, where $L = \{\ell_0, \ell_1, \dots, \ell_n\}$ is a set of locations representing program counter values, $E = L \times \mathit{Ops} \times L$ is a set of directed edges representing possible control flow paths. The edges are labeled with operations in $\mathit{Ops}$ which get executed when control jumps from one location to another. The distinguished initial location $\ell_0 \in L$ marks the entry point of the program and the special error location $\ell_e \in L$ embodies an undesirable state in the program. In our work, we use $\ell_e$ to represent failing assertions in a C program. We focus on \emph{assertion checking}, i.e., showing that $\ell_e$ cannot be reached in any feasible executions of the program.

\paragraph{CEGAR.} A typical drawback of using model checking techniques is their high computational complexity. Abstraction-based methods are often applied to overcome this issue. However, it is not straightforward to find the proper precision of abstraction that is coarse enough to reduce complexity but fine enough to prove the desired property. Counterexample-Guided Abstraction Refinement (CEGAR) is an automatic algorithm that starts with a coarse initial abstraction and refines it based on the counterexamples until the proper precision is reached~\cite{clarke03}. CEGAR was first described using transition systems and predicate abstraction, but since then many variants have been developed.

In this work we use our generic CEGAR framework~\cite{forte2016} for verification, which incorporates many variants of the algorithm. The framework explores the abstract state space in a given \emph{abstract domain} with a given \emph{exploration strategy}. Currently, predicate~\cite{graf97} and explicit value~\cite{clarke04} domains are supported with breadth- and depth-first search exploration strategies. The abstract state space is an over-approximation of the original, therefore if no erroneous abstract state is reachable, the original model is also correct. However, if an abstract counterexample is found, its feasibility in the original model is checked. A feasible counterexample corresponds to a failure in the original model. On the other hand, if the counterexample is infeasible (also called spurious), the abstraction is refined. The framework currently supports refinement based on binary or sequence interpolation~\cite{mcmillan05, vizel09} and unsat cores~\cite{leucker15}. The model is then checked again with the refined abstraction and this procedure is repeated until the model is proved to be safe or a feasible counterexample is found. We first defined our framework for transition systems~\cite{forte2016} but since, the algorithms have been adapted to CFAs.

\subsection{Program Representations}
In this paper, we focus on the optimization and verification of C programs. In order to do this, we transform the textual representation into structures more sufficient for dependency analysis, which are needed for the size reduction transformations.

While the control flow automaton representation of a program is suitable for verification, most analysis and transformation algorithms are defined over the language-agnostic formalism of \emph{control flow graphs} (CFG)~\cite{DragonBook}. A CFG is a 4-tuple $(S,E, s_0, s_q)$, where $S = \{s_0, s_1, \dots, s_n\}$ is a set of atomic instructions and $E = \{ (s_i, s_j), (s_k, s_l), \dots \}$ is a set of directed edges representing possible control flow paths. The edge $(s_i, s_j) \in E$ iff there is a conditional or unconditional jump from $s_i$ to $s_j$ in the program.
The distinguished $s_0 \in S$ and $s_q \in S$ nodes mark the entry and exit points of the program, respectively. As there is a one-to-one correspondence between CFG nodes and program instructions, we shall use the two terms interchangeably.

During analysis, it is often useful to know whether an instruction writes variables that are later read by another instruction. We say that instruction $s$ \textit{defines} the variable $v$ if $s$ assigns a value to $v$. A \textit{definition} is a pair $d=(s, v)$ where $s$ is an instruction and $v$ is the variable defined in $s$.  Given a node $t \neq s$, $d$ is a \textit{reaching definition} for $t$, if there is a control flow path between $s$ and $t$, which contains no other definition of $v$. If $s$ has a reaching definition for $t$, then $t$ is said to be \textit{flow dependent} on $s$. A common alternative name for flow dependency is \textit{data dependency}.

This flow dependency information can be stored in a structure known as use-define chain (UD-chain for short)~\cite{DragonBook}. Given a program $P$ with the definitions $D = \{d_1, d_2, \dots, d_n\}$ and an instruction set $S = \{s_1, s_2, \dots, s_k\}$, the \textit{use-define chain} of $P$ is a set of pairs $\{ (s_1, D_1), (s_2, D_2), \dots, (s_k, D_k) \}$ where $D_i \subseteq D$ is the set of definitions reaching $s_i$ for all $1 \le i \le k$. Given an instruction $s$, querying the UD-chain yields all instructions on which $s$ flow depends.

It is also useful to know which instructions decide whether some other instruction gets executed or not. This information can be obtained by analyzing the post-dominance relations of the program~\cite{Ferrante87}. A node $s$ \textit{post-dominates} a node $t$, if every control flow path between $t$ and $s_q$ (the exit location) contains $s$. If $s \neq t$, then $s$ \textit{strictly post-dominates} $t$. A node~$t$ is \textit{control dependent} on a node~$s$ if there is a path from $s$ to $t$ where all nodes are post-dominated by $t$ and $t$ does not strictly post-dominate $s$.

In order to represent these dependency relations of a program, we use a structure known as \textit{program dependence graph} (PDG)~\cite{Ferrante87}. A PDG is a triple $(S, C, D)$, where $S = \{s_1, \dots, s_n\}$ is a set of instructions, $C$ is a set of control dependency edges, and $D$ is a set of data dependency edges. The edge $(s_i, s_j) \in C$ if $s_j$ control depends on $s_i$. The edge $(s_k, s_l) \in D$ if $s_l$ flow depends on $s_k$.

\subsection{Compiler Optimizations}
\label{sec:background:compileropt}
In order to reduce the resulting model's size and complexity, we use \textit{optimization transformations} usually known from compiler theory~\cite{DragonBook}. \textit{Constant folding} finds and evaluates constant expressions in the program during compile time. \textit{Constant propagation} substitutes variables having a constant value with their respective constant literal. As constants can be local or global, both information need to be propagated. This can be achieved by querying the use-define information of the program. In some cases, constant propagation and folding are able to replace branching decisions with the literals \texttt{true} or \texttt{false}. \textit{Dead branch elimination} examines these branches and removes inviable execution paths (e.g.\ the \texttt{true} path of a branch with a condition of the boolean literal \texttt{false}). 

For easing interprocedural analysis, we also use a technique known as \textit{function inlining}. Function inlining is the procedure of replacing a function call with the callee’s body.
In our work, we use it to obtain more information on the behavior of an interprocedural program, as without more thorough interprocedural analysis, function calls would act as black boxes. A model checker algorithm may extract more information from the entire inlined function body than from merely just a function call.

\subsection{Program Slicing}
\label{sec:background:slicing}
Program slicing is a size reduction technique that attempts to remove all nodes irrelevant to a given instruction and a given set of variables~\cite{Weiser81}.
Let $P$ be an input program, let $V$ be a subset of its variables, and let $s$ be an instruction in $P$.
A \textit{program slice} $P'$ of $P$ \textit{with respect to the criterion} $(s, V)$ is a subprogram of $P$, which produces the same output and assigns the same values to $V$ as the original program $P$ in its statement $s$.

Figure~\ref{fig:slice-ex:a} presents an example with a simple C code snippet, which
calculates the sum of the natural numbers less than 11, and asserts that the loop counter and the calculated sum cannot be zero. A slice of this program with respect to the criterion $(9, \{i\})$ is shown in Figure~\ref{fig:slice-ex:b}. As it can be observed, the only statements preserved are those relevant to the criterion instruction.

\begin{figure}[!htb]
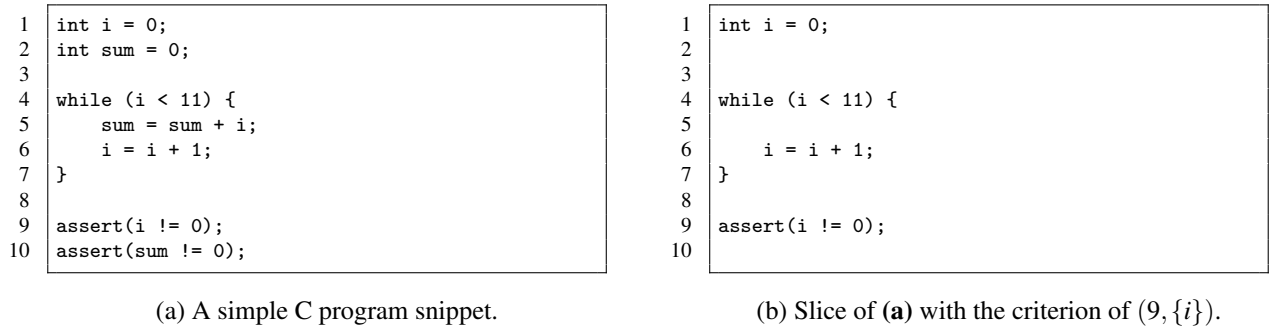

	\centering
	\begin{subfigure}[b]{.45\linewidth}
		\begin{lstlisting}
int i = 0;
int sum = 0;

while (i < 11) {
	sum = sum + i;
	i = i + 1;
}

assert(i != 0);
assert(sum != 0);\end{lstlisting}
		\caption{A simple C program snippet.}
		\label{fig:slice-ex-c:a}
	\end{subfigure}
\hfill
	\begin{subfigure}[b]{.45\linewidth}
		\begin{lstlisting}
int i = 0;


while (i < 11) {

	i = i + 1;
}

assert(i != 0);
 \end{lstlisting}
		\caption{Slice of \textbf{(a)} with the criterion of $(9, \{i\})$.}
		\label{fig:slice-ex-c:b}
	\end{subfigure}
	\caption{An example on slicing.}
	\label{fig:slice-ex-c}
\end{figure}

As we are using verification for detecting failing assertions, we use program slicing to split a larger input program into several smaller chunks, with the criteria being the assertion instructions and the variables they use. This will result in smaller verifiable programs instead of a larger one, more precisely every assertion of the program gets verified independently. A proof of a slice's faultiness also indicates that the original program is faulty. On the other hand, if all slices are safe, it means that no assertion in the original program can fail, meaning that the whole program is also safe.

Several program slicing techniques exist, depending on the use case (model checking, debugging, test generation, etc.). In our current work, we use three methods, all of which are suitable or especially tailored for model checking.

\paragraph{Backward slicing.} The most commonly used approach for slicing is a technique known as \textit{backward slicing}. Backward slicing produces accurate slices, while retaining all instructions which are crucial to the slicing criterion. The size of the resulting slices are roughly around 30\% of the size of the original program on average~\cite{Binkley07}. Given a criterion instruction $s$, backward slicing finds all nodes on which $s$ data depends transitively. As some branching decisions may affect whether a particular instruction is reachable or not, these also need to be included in the slice. These branching decisions are the same as the control dependencies of a given instruction. Of course, these dependencies can have dependencies which need to be taken into account. Therefore backward slicing is done by retaining all instructions on which the criterion control or flow depends transitively. This information can be queried from a program dependence graph.

\begin{figure}[!htb]
	\centering
	\includegraphics[width=.6\linewidth, keepaspectratio]{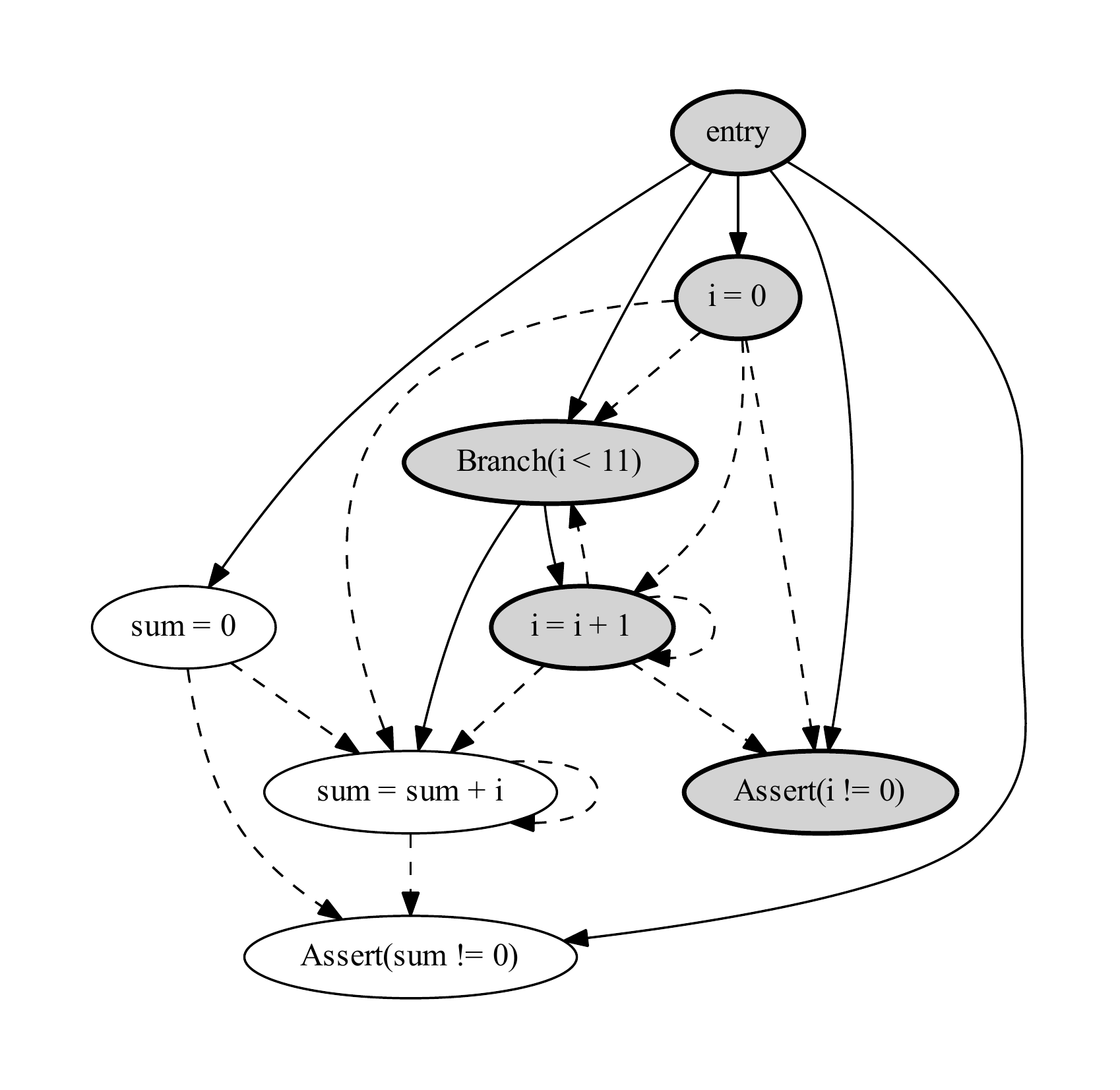}
	\caption{Slicing on the \texttt{Assert(i!=0)} node using a program dependence graph.}
	\label{fig:while-slice-ex-pdg}
\end{figure}

A backward slicer algorithm marks all nodes which are backwards reachable (walking backwards on both data and control dependency edges) from the criterion node in the program dependence graph~\cite{Ferrante87}. As the PDG explicitly shows the control and flow dependency relations of every instruction, this method will include all required nodes in the slice. Figure~\ref{fig:while-slice-ex-pdg} shows an example of this procedure (for the same code as in Figure~\ref{fig:slice-ex-c}) with the assertion node being the criterion. Solid lines represent control dependency, dashed lines represent flow dependency, and filled nodes are those which are backwards reachable from the criterion node.

While backward slicing is simple and accurate, there are other algorithms that sacrifice faithfulness for even greater program size reduction. Such algorithms can be rather useful in the context of model checking. As discussed above, given a slicing criterion $\gamma = (s, V)$, backward slicing attempts to find all transitive control and flow dependencies of $\gamma$. In many cases the control dependencies are only required to keep the structure of the program, their respective branch decisions do not affect $\gamma$. This idea serves as the basis of the slicing techniques described below.

\paragraph{Thin slicing.} A technique known as \textit{thin slicing}~\cite{Sridharan07} aggressively reduces the program size by retaining flow dependencies only. All control dependencies are replaced with nondeterministic boolean predicates (called \textit{abstract predicates} -- we denote such predicates with $\phi$). After performing this substitution, the thin slicing algorithm behaves the same as backward slicing (in the sense that we are performing a backward search in the PDG).
Abstract predicates have no flow dependencies, therefore no such dependencies will be included in the slice, thus allowing a much larger reduction.

On the other hand, abstract predicates may introduce new possible execution paths into the program.
In the original program (or a backward slice), a verification algorithm may realize that a certain execution path is inviable due to the value of a branch condition. However, if the program has abstract predicates only, this assumption cannot be made. This also means that the verifier algorithm has to discover all program paths in the resulting slice.

The inclusion of dynamically inviable execution paths also means that verification algorithms may produce more spurious counterexamples on thin slices. Finding whether a counterexample can hold for the original program requires the refinement of the thin slice (which means the inclusion of previously abstracted control dependencies), and running the verifier again on the refined slice. The details of such refinements will be discussed later on.

\paragraph{Value slicing.} A possible middle ground between thin and backward slices is called \textit{value slicing}~\cite{Kumar15}. Value slicing recognizes which control dependencies actually determine the value of the criterion variables and retains those as well. This results in slightly larger program slices compared to thin slicing. However, verification algorithms produce less spurious counterexamples on these slices, thus they require less refinement. Value slicing introduces the term \textit{value-impacting instruction}. Given a slicing criterion $\gamma$, an instruction $s$ value-impacts $\gamma$ if any of the following conditions hold:
\begin{enumerate}
	\item The slicing criterion $\gamma$ flow depends on $s$.
	\item There exists an instruction $t$ which value-impacts $\gamma$ and $t$ flow depends on $s$.
	\item The instruction $s$ is branch decision with conditional jumps to the instructions $t_1$ and $t_2$, from which there exist paths $\pi_1 = \{t_1, \dots, \gamma\}$ and $\pi_2 = \{t_2, \dots, \gamma\}$ in the CFG,
	starting with $t_1$, $t_2$ and ending in the first occurrence of $\gamma$. Furthermore, there exists a node $t \neq s$ such that $t$ value-impacts $\gamma$ and $t$ is the first value-impacting node along $\pi_1$ and $t$ is not the first value-impacting node along $\pi_2$.
\end{enumerate}

The first and second conditions make sure that the flow dependencies of the remaining instructions are all accounted for.
The third condition marks branch decisions which will be retained.
A branch decision is value-impacting (and therefore should not be abstracted) if both its outgoing edges are program paths to the slicing criterion and it decides whether a value-impacting instructions gets executed or not.

As mentioned before, value and thin slicing may require refinement in order to prove that a counterexample encountered is not a spurious one.
This can be done by selecting an abstract predicate $\phi$ from the slice and adding it to the slicing criterion, then performing a slicer algorithm again. The selection of such abstract predicates can be done randomly (our current implementation does this) or by using certain heuristics (e.g. distance from the criterion node in the PDG, etc.).
Note that the slicing algorithm used during the refinement can be different than the initial slicer.
For example, using thin slicing as refinement means that we are refining the slice with a single abstract predicate each time.
On the other hand, using value slicing as refinement means that the refined slice may include multiple abstract predicates at once.

\begin{figure}[!htb]
	\begin{subfigure}[b]{.28\linewidth}
		\begin{lstlisting}
extern int fn1();
extern int fn2();
extern int fn3(int x, int y);

int main() {
	int i = 0, j = 0;	
	int t = fn1();
	int x = fn3(i, j);
	int y = 0;
	
	while (t < 1000) {
		int s = fn2();		
		if (s == 1) {
			y = y + x;
		} else {
			i = i + 1;
			j = j + 1;
		}
		
		assert(y != 0);
		
		x = fn3(i, j);
		t = fn1();
	}
	
	return 0;
}\end{lstlisting}
		\caption{Backward slice.}
		\label{fig:slice-ex:a}
	\end{subfigure}
	\hfill
	\begin{subfigure}[b]{.28\linewidth}
		\begin{lstlisting}[mathescape=true]
// fn1() removed
extern int fn2();
extern int fn3(int x, int y);

int main() {
	int i = 0, j = 0;	
	
	int x = fn3(i, j);
	int y = 0;
	
	while ($\phi$) {
		int s = fn2();		
		if (s == 1) {
			y = y + x;
		} else {
			i = i + 1;
			j = j + 1;
		}
		
		assert(y != 0);
		
		x = fn3(i, j);
		
	}
	
	return 0;
}\end{lstlisting}
		\caption{Value slice.}
		\label{fig:slice-ex:b}
	\end{subfigure}
	\hfill
	\begin{subfigure}[b]{.28\linewidth}
		\begin{lstlisting}[mathescape=true]
// fn1() removed
// fn2() removed
extern int fn3(int x, int y);

int main() {
	int i = 0, j = 0;	
	
	int x = fn3(i, j);
	int y = 0;
	
	while ($\phi_1$) {
		
		if ($\phi_2$) {
			y = y + x;
		} else {
			i = i + 1;
			j = j + 1;
		}
		
		assert(y != 0);
		
		x = fn3(i, j);

	}

	return 0;
}\end{lstlisting}
		\caption{Thin slice.}
		\label{fig:slice-ex:c}
	\end{subfigure}
	\caption{An example on backward, value, and thin slices with the criterion $(20, \{y\})$.}
	\label{fig:slice-ex}
\end{figure}

Figure~\ref{fig:slice-ex} offers an example (a simplified version of an example in \cite{Kumar15}) on the different slicing methods described above. Suppose \texttt{fn1}, \texttt{fn2}, and \texttt{fn3} are complex operations with multiple different computations and all examples are slices on the criterion $(20, \{y\})$.
As it can be seen in Figure~\ref{fig:slice-ex:a}, a backward slice retains all instructions which are relevant to the given criterion in any way. The thin slice (shown in Figure~\ref{fig:slice-ex:c}) abstracted branch conditions away, by replacing them with $\phi_1, \phi_2$. Note that because of the removed branch conditions in line $11$ and $13$, the calculation of their used variables is also omitted from the slice. This also allowed the removal of \texttt{fn1} and \texttt{fn2}.
The value slice (Figure~\ref{fig:slice-ex:b}) also abstracted the loop condition in line $11$ away, but it kept the condition in line $13$. This is because of the fact that the value of $y$ is actually determined here:
line $13$ decides whether the value-impacting assignment in line $14$ gets executed or not. If the branch condition is true, $y$'s value changes, but if it is false, $y$ stays the same. The function \texttt{fn2} was retained as well, because line $13$ depends on the value of $s$, which was defined in line $12$ by using the value returned by \texttt{fn2}.

\section{Transformation Workflow}
\label{sec:workflow}

In this section we propose a workflow that is able to transform C programs to control flow automata. In order to reduce the size of the resulting models, we enhance this process with compiler optimizations and program slicing. The resulting models can then be verified using an arbitrary verification algorithm. In this paper, we use CEGAR-based algorithms~\cite{forte2016} for this purpose.

\begin{figure}[!htb]
	\centering
	\begin{tikzpicture}
\tikzstyle{box}=[rectangle,align=center,draw, thick,minimum width=2.3cm,minimum height=0.6cm,rounded corners]

\node[box] (cco) at (0,0) {C code};
\node[box] (ast) at (3.5,0) {AST};
\node[box] (cfg) at (7,0) {CFG};
\node[box] (cfa) at (10.5,0) {CFA};
\node[box] (dt) at (4.5,-1.5) {Dominators};
\node[box] (udc) at (7,-1.5) {UD-chains};
\node[box] (pdg) at (5.75,-3) {PDG};
\node[box] (cg) at (9.5,-1.5) {Call Graphs};

\draw[-angle 45,very thick](cco)--(ast) node[midway,above]{};
\draw[-angle 45,very thick](ast)--(cfg) node[midway,above]{};
\draw[-angle 45,very thick](cfg)--(cfa) node[midway,above]{};
\draw[-angle 45,very thick] (cfg)to[out=70,in=110,looseness=10](cfg);

\draw[-angle 45,very thick,dashed](cfg)--(dt) node[midway,above]{};
\draw[-angle 45,very thick,dashed](cfg)--(udc) node[midway,above]{};
\draw[-angle 45,very thick,dashed](dt)--(pdg) node[midway,above]{};
\draw[-angle 45,very thick,dashed](udc)--(pdg) node[midway,above]{};
\draw[-angle 45,very thick,dashed](cfg)--(cg) node[midway,above]{};

\end{tikzpicture}
	
	\caption{Transformation workflow.}
	\label{fig:workflow-overview}
\end{figure}
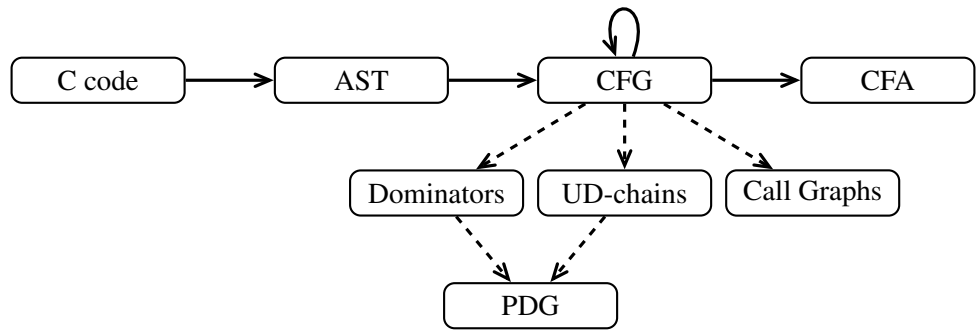

An overview of the transformation workflow is shown in Figure~\ref{fig:workflow-overview}.
First, we take a C source code file as an input. This source code is then parsed into an \textit{abstract syntax tree} (AST), which describes the syntactic structure of the program. The abstract syntax tree is then transformed into a control flow graph. This CFG is then simplified by the application of compiler optimizations, namely function inlining, constant propagation, constant folding, and dead branch elimination. For further size reduction, we also apply program slicing. Several dependency structures are required for these transformations:
\begin{itemize}
	\item program dependence graphs for slicing,
	\item dominator relation information for constructing the PDG,
	\item \textit{use-define} information (\textit{UD-chains}) for the PDG and also for constant propagation,
	\item and \textit{call graphs} for function inlining.
\end{itemize}

After running the optimization passes, we perform slicing on the resulting CFG. This operation splits a single CFG into several smaller ones. Our framework supports all backward, value, and thin slicing. Currently the slicer criteria are the assertion instructions (calls to the \texttt{assert} function in C) and their associated variables in the control flow graph, meaning that each assertion gets its own slice. These slices are then transformed into control flow automata. Due to the slicing criteria, error locations in the resulting automata represent failing assertions of the original program.

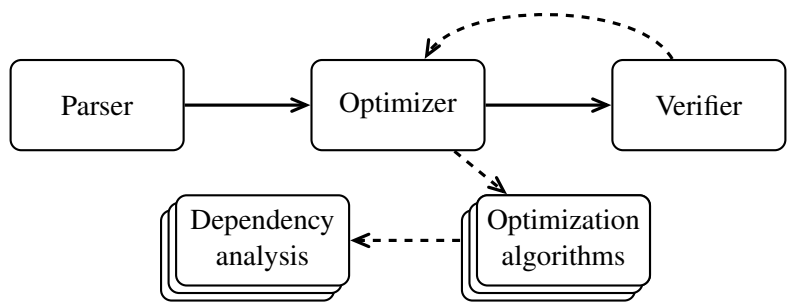
\begin{figure}[!htb]
	\centering
	\begin{tikzpicture}
	\tikzstyle{box}=[rectangle,align=center,draw,thick,minimum width=2.3cm,minimum height=1.2cm,rounded corners,fill=white]
	
	\node[box] (parser) at (3,0) {Parser};
	\node[box] (opt) at (7,0) {Optimizer};
	\node[box] (bmc) at (11,0) {Verifier};
	
	\node[box] (deps0) at (5, -2) {};
	\node[box] (deps1) at (5.1, -1.9) {};
	\node[box] (deps2) at (5.2, -1.8) {Dependency\\ analysis};
	
	\node[box] (optalgsX) at (9.2, -1.8) {};
	\draw[-angle 45,very thick,dashed](optalgsX)--(deps2);
	\node[box] (optalgs0) at (9, -2) {};
	\node[box] (optalgs1) at (9.1, -1.9) {};
	\node[box] (optalgs2) at (9.2, -1.8) {Optimization\\ algorithms};
	
	\draw[-angle 45,very thick,dashed](opt)--(optalgs2);
	\draw[-angle 45,very thick](parser)--(opt);
	\draw[-angle 45,very thick](opt)--(bmc);
	\draw[-angle 45,very thick, dashed](bmc)to[out=120,in=60,looseness=0.7](opt);

	\end{tikzpicture}
	
	\caption{Architecture of the implementation.}
	\label{fig:impl-architecture}
\end{figure}

\paragraph{Implementation.}
We implemented a prototype of our workflow in Java. Figure~\ref{fig:impl-architecture} offers an overview of the overall architecture, which builds on three loosely dependent components: parser, optimizer and verifier.

The parser component uses the parsing library of the Eclipse C/C++ Development Tools plug-in (CDT).\footnote{\url{http://www.eclipse.org/cdt/}} The CDT parser handles source code lexing and parsing, after which it yields an abstract syntax tree. This AST is then transformed into a control flow graph. At the current state of the project, only a restricted set of C language features is supported. The current implementation only allows the usage of control structures (\textbf{if-then-else}, \textbf{do-while}, \textbf{while-do}, \textbf{switch}, \textbf{break}, \textbf{continue}, \textbf{goto}) and non-recursive functions. Types are restricted to integers and Booleans only. Arrays and pointers are not supported at the moment. 

The optimizer module performs the optimization transformations. This component is able to run a configurable number of optimization passes. The implementation currently supports constant propagation, constant folding, dead branch elimination, and function inlining. Some optimizations may require certain dependency information of the input program, therefore they need to be able to query this information at any time. The optimizer also handles program slicing. The module is able to use an arbitrary slicing algorithm (which conforms to a given interface per se) to perform slicing. Currently we support the methods presented in the previous sections, namely traditional backward slicing, thin slicing, and value slicing. After the acquisition of the smaller program slices from the slicer algorithm, the optimizer transforms these CFG slices into control flow automata.

These automata are then checked by the verifier component for assertion violations. While this component is modular and replaceable, our current implementation focuses solely on predicate abstraction in conjunction with a CEGAR-based algorithm~\cite{forte2016} adapted to CFAs. As refinement may be needed due to the presence of value and thin slicing, it is possible that the optimizer has to run again and refine the slice.

\section{Evaluation}
\label{sec:evaluation}

In this section, we evaluate and compare the effects of optimizations regarding the size of the model and the efficiency of verification. First we present the programs used and the environment, and then we discuss the results.

\subsection{Objects and Environment}
We performed our evaluation on programs from the Competition~on~Software Verification~(SV-Comp)~\cite{Beyer16} repertoire. This competition aims to compare the soundness and performance of software verification tools. In our evaluation we use some of its tasks to compare and evaluate the performance of the different algorithms implemented in our tool. Other tools are currently not considered as we only aim to evaluate the different algorithms within our framework. The verification tasks used in our work can be separated into three main categories, which are described below.

\begin{description}
	\item [locks] This task set consists of small (100-150 LOC) locking mechanisms described with nondeterministic integers and \textbf{if-then-else} constructs.
	This category consists of files, which have several assertion statements within, therefore producing many (10-15) small slices.
	\item [eca] The ECA (event-condition-action) category describes large (500-600 LOC) event-driven reactive systems. The events are represented with nondeterministic variables.
	The files in this category are special in a sense that one large problem is already split up into several files, with each file containing a single assertion, hence a single slice.
	\item [ssh-simplified] These tasks describe large (500-600 LOC) server-client systems. While these systems are rather complex, verifying the server-client communication is not part of this task, such factors are abstracted away with nondeterministic variables. These programs also produce a single slice.
\end{description}

As our framework currently supports slicing on \texttt{assert} instructions, we used slightly modified versions of the programs listed above. The original competition specifications considered a program faulty if a particular error function was called. We replaced such calls with an \texttt{assert(false)} instruction so that the call of the error function will lead to the special error location.

Variables of the evaluation are listed in Table~\ref{tab:vars} grouped into three main categories: parameters of the program (input), parameters of the algorithm configuration (input), metrics of the algorithm (output).
In the figures configurations are given by the abbreviation of the possible input options.
For example, \varvalue{VFD} denotes value slicing (\varvalue{V}), without optimizations (\varvalue{F}, false), with DFS search strategy (\varvalue{D}). Individual slices are referred to by concatenating the file name and the slice number with a hash sign (e.g., \texttt{locks/locks11\_true.c\#2})

\begin{table*}[!htb]
	\caption{Variables of the experiment.}
	\centering
	\begin{tabularx}{\linewidth}{|l|l|l|X|}
		\hline
		Category & Name & Type & Description\\\hhline{|=|=|=|=|}
		\multirow{4}{1.1cm}{Input\\(program)} & \varname{File} & String & Unique name (and path) of the instance. \\\cline{2-4}
		& \varname{Slice No.} & Integer & Index of the slice (assertion) being verified. If the program is not sliced with a given configuration, this variable is ignored. \\\hhline{|=|=|=|=|}
		
		\multirow{4}{1.1cm}{Input (config.)} & \varname{Slicer} & Factor & Slicing algorithm. Possible values: \varvalue{NONE}, \varvalue{THIN}, \varvalue{VALUE}, \varvalue{BACKWARD}.\\\cline{2-4}
		& \varname{Optimizations} & Boolean & Indicates if constant propagation and dead branch elimination transformations are used.\\\cline{2-4}
		& \varname{Search} & Factor & Search strategy during verification. Possible values: \varvalue{BFS}, \varvalue{DFS} (breadth- and depth-first search).\\\hhline{|=|=|=|=|}
		
		\multirow{5}{1.1cm}{Output (metrics)} & \varname{Safe} & Boolean & Verification result, indicates whether the given slice was deemed safe or unsafe by the verifier.\\\cline{2-4}
		& \varname{InitLocs} & Integer & Initial location count in the CFA corresponding to the slice.\\\cline{2-4}
		& \varname{InitEdges} & Integer & Initial edge count in the CFA corresponding to the slice.\\\cline{2-4}
		& \varname{ArgSize} & Integer & Number of nodes in the Abstract Reachability Graph (ARG), i.e., the number of explored abstract states.\\\cline{2-4}
		& \varname{EndLocs} & Integer & Final location count in the CFA corresponding to the slice.\\\cline{2-4}
		& \varname{EndEdges} & Integer & Final edge count in the CFA corresponding to the slice.\\\cline{2-4}
		& \varname{Optimization time} & Integer & Execution time of the optimizer component (including optimizations, slicing, and slice refinements), in milliseconds.\\\cline{2-4}
		& \varname{Verification time} & Integer & Execution time of the verification algorithm, in milliseconds. \\\hline
		
	\end{tabularx}
\label{tab:vars}
	
\end{table*}

All measurements were executed on a 64-bit Windows~7 virtual machine with 2 cores ($2.50$~GHz) and 16~GB~RAM. Each slice was tested with the timeout of 3 minutes.
To avoid interprocedural analysis, function inlining was applied to every input program, regardless of the configuration. In our experiment, using a thin slicer for refinements was deemed rather unfavorable, because it runs a new iteration for every abstract predicate. In order mitigate this, we use value slicing as the refinement algorithm for both the thin and value slicer.

\subsection{Results and Discussion}

In our experiment, we evaluated 9 input programs (3 from \textbf{locks}, 4 from \textbf{eca} and 2 from \textbf{ssh}). With slicing, these programs have produced 50 slices.
We checked these slices individually and also checked the original programs without slicing. Given that we have $3$ slicing algorithms ($+1$ for no slicing), $2$ search strategies and $2$
possible values (\varvalue{true} or \varvalue{false}) regarding compiler optimizations, the total number of configurations is $16$. This gives us a total number of $4 \cdot 9 + 12 \cdot 50 = 636 $ measurements. For convenience, from this point forward, we refer to any input model (a whole, unsliced program, or a specific slice) as ''slice''. The number of successful executions (no timeout) is $570$, the number of safe results is $484$.

\begin{figure}[!htb]
	\centering
	\includegraphics[width=\linewidth]{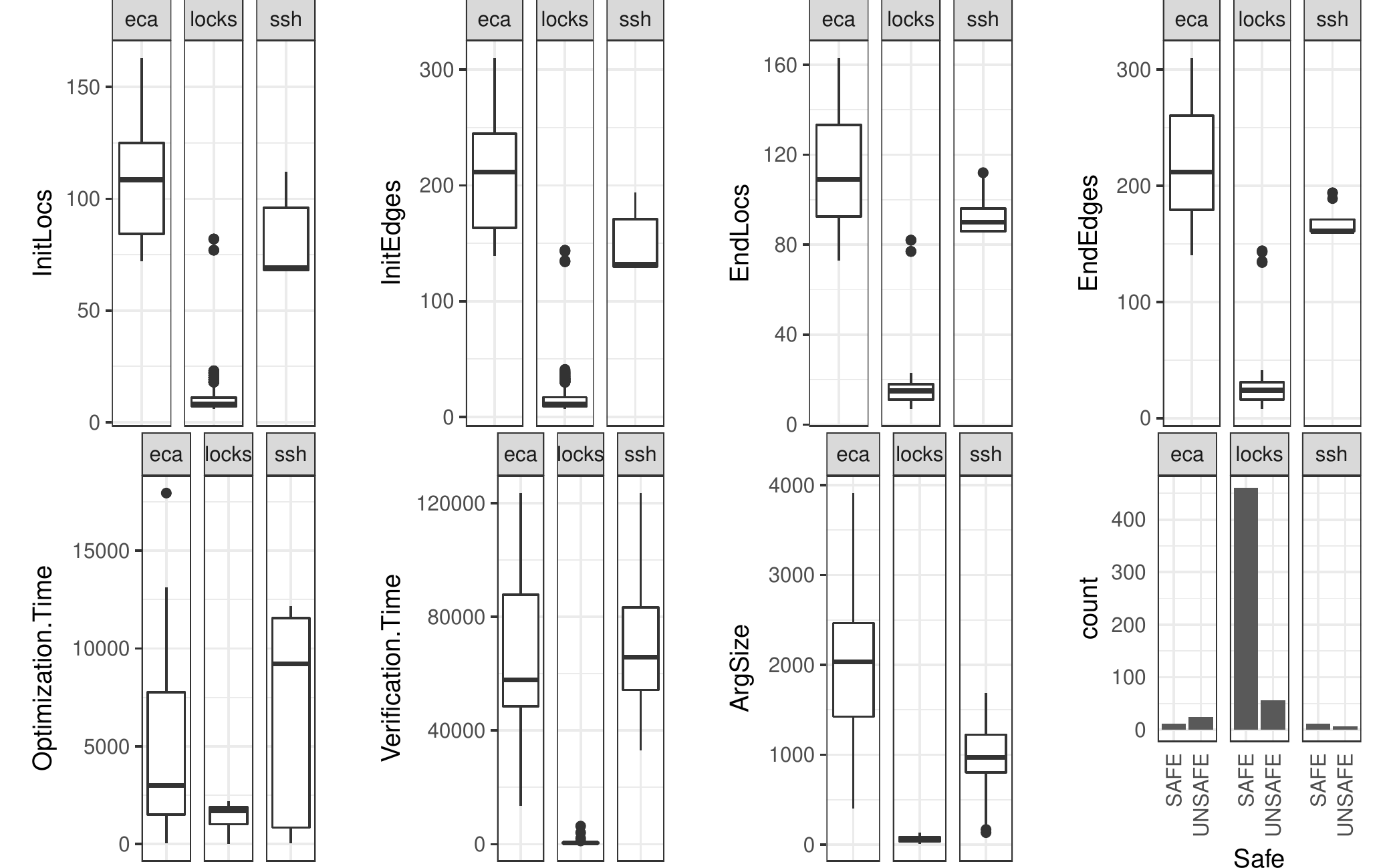}
	\caption{Overview of output variables.}
	\label{fig:out-vars-categories}
\end{figure}

Figure~\ref{fig:out-vars-categories} gives a high level overview of the distribution and range of output variables, grouped by task categories. As it can be seen, values for the \textbf{eca} and \textbf{ssh} task sets are on a rather similar scale. The \textbf{locks} set, however, produces significantly smaller numbers and more outliers. Therefore, we present those results separately from the other two categories, where this is required. We also found that there is a strong correlation between the location and edge count of the CFAs (with an $R^2$ value of~$0.998$), therefore we only use the former to describe the size of a CFA.

\subsubsection{Impact of Slicing and Optimizations on CFA Size}
\begin{figure}[!htb]
	\centering
	\begin{subfigure}[b]{0.49\linewidth}		
		\includegraphics[width=\linewidth]{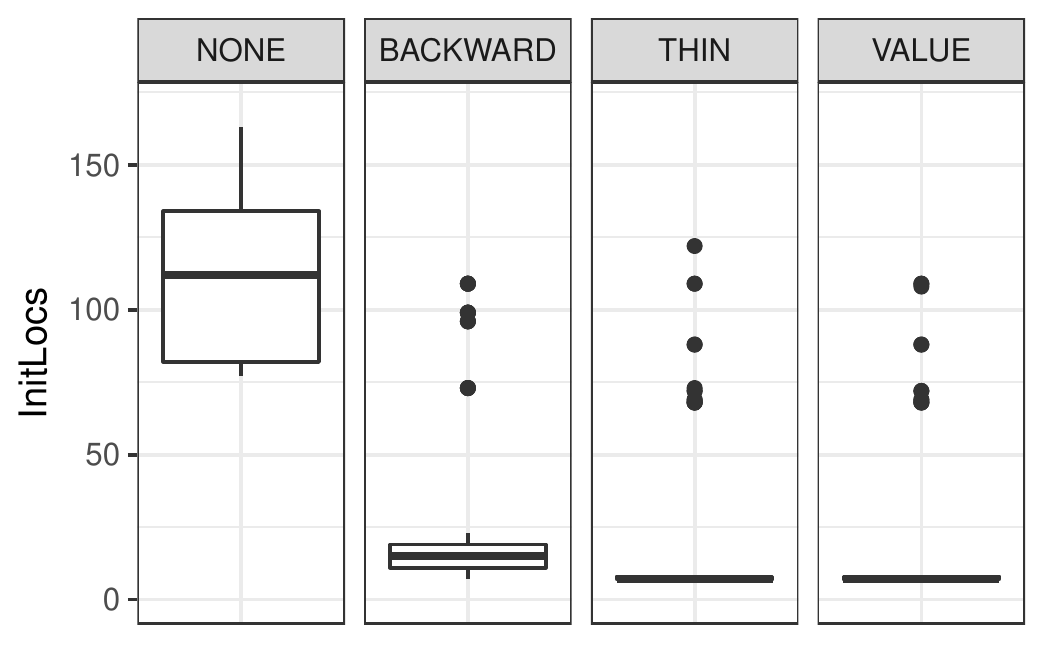}
	\end{subfigure}
	\begin{subfigure}[b]{0.49\linewidth}		
		\includegraphics[width=\linewidth]{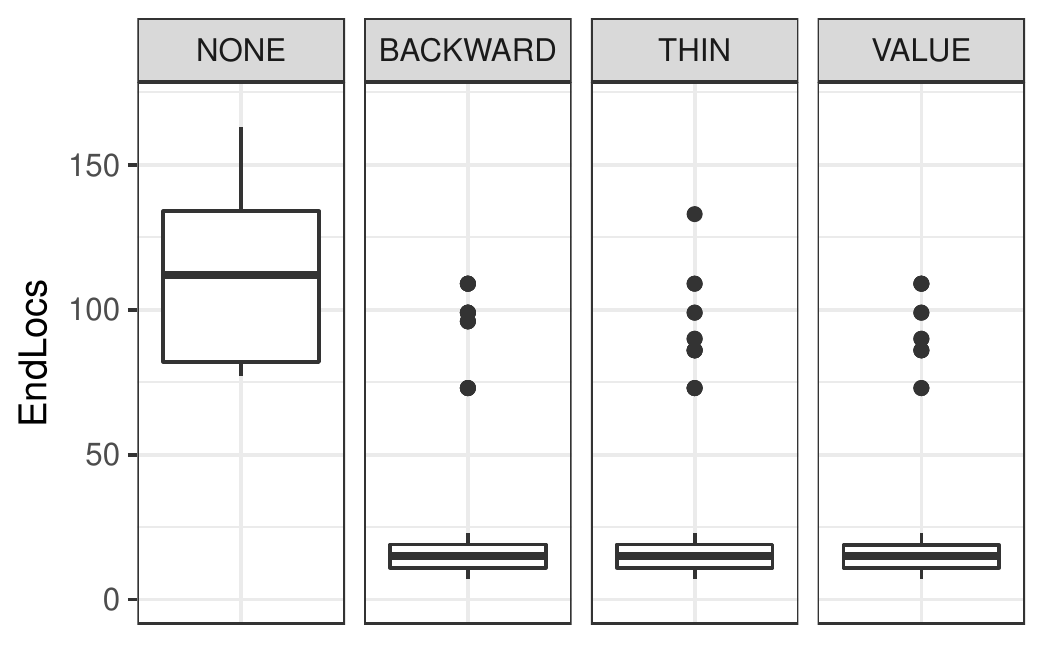}
	\end{subfigure}
	
	\caption{Comparison of the effect of different slicers on the initial and final CFA size.}
	\label{fig:slicer-cfa-size-box}
\end{figure}

An overview of the size reduction effect of the different slicing methods on the initial and final size of the CFA is shown in Figure~\ref{fig:slicer-cfa-size-box}.
It can be seen that all three slicing techniques reduce the size of most programs significantly compared to the case when no slicing is applied. As it can be seen in the left, thin and value slicing allow even greater reduction initially, but the plots on the right show that the size of the final automata (after refinements) becomes roughly the same as in the case of backward slicing.

\begin{figure}[!htb]
	\centering
	\includegraphics[width=0.6\linewidth]{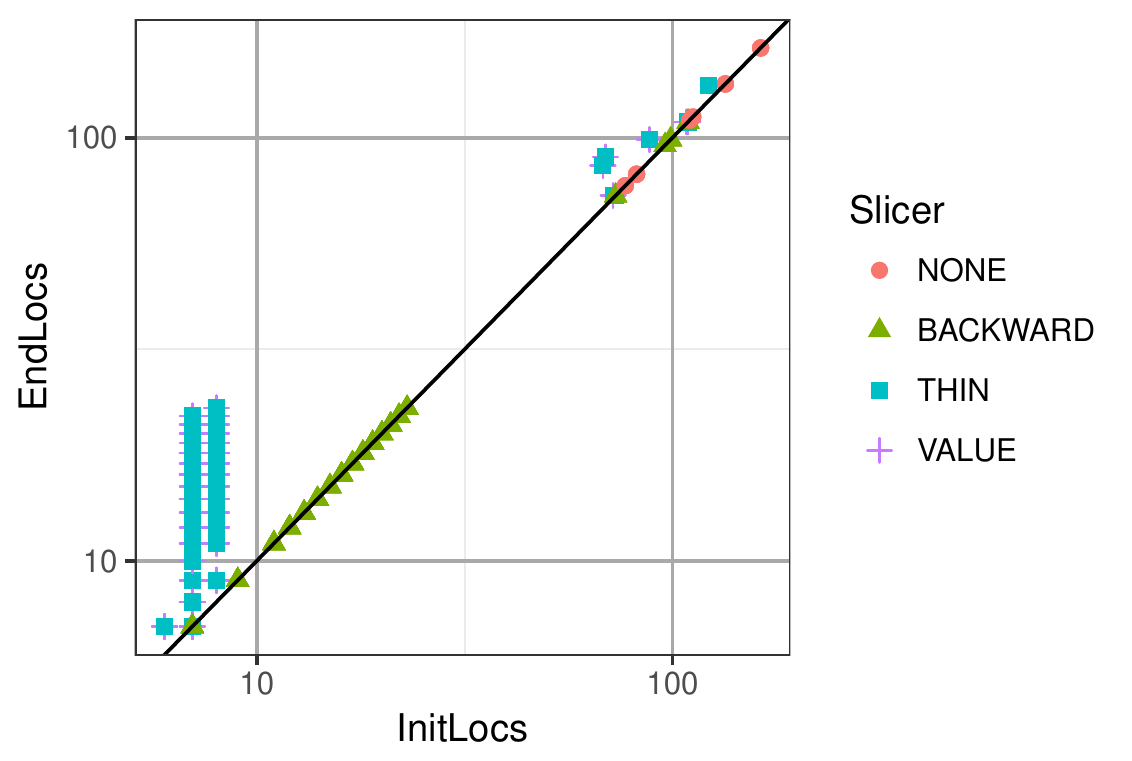}
	\caption{Overview of the initial and final CFA size with different slicers.}
	\label{fig:slicer-cfa-size}
\end{figure}

Figure~\ref{fig:slicer-cfa-size} compares the initial and final size of the CFA for each slice individually.
Naturally, no difference can occur in initial and final size if we use no slicing.
As backward slicing requires no refinement, size of the backward slices also stays the same.
In contrast, value and thin slicing usually requires refinements, which iteratively increase the size of the CFA.

Curiously, the measurements of the value slicer are same as the thin slicer's.
This is due to the fact that the value slicer only retains branches that make a choice between the possible values of a variable required by the criterion node.
The SV-Comp task set contains programs with a distinctive artificial structure and no such branches are present in the given programs,
therefore the heuristics used by the value slicing algorithm could not work in this case.

\begin{figure}[!htb]
	\centering
	\includegraphics[width=0.6\linewidth]{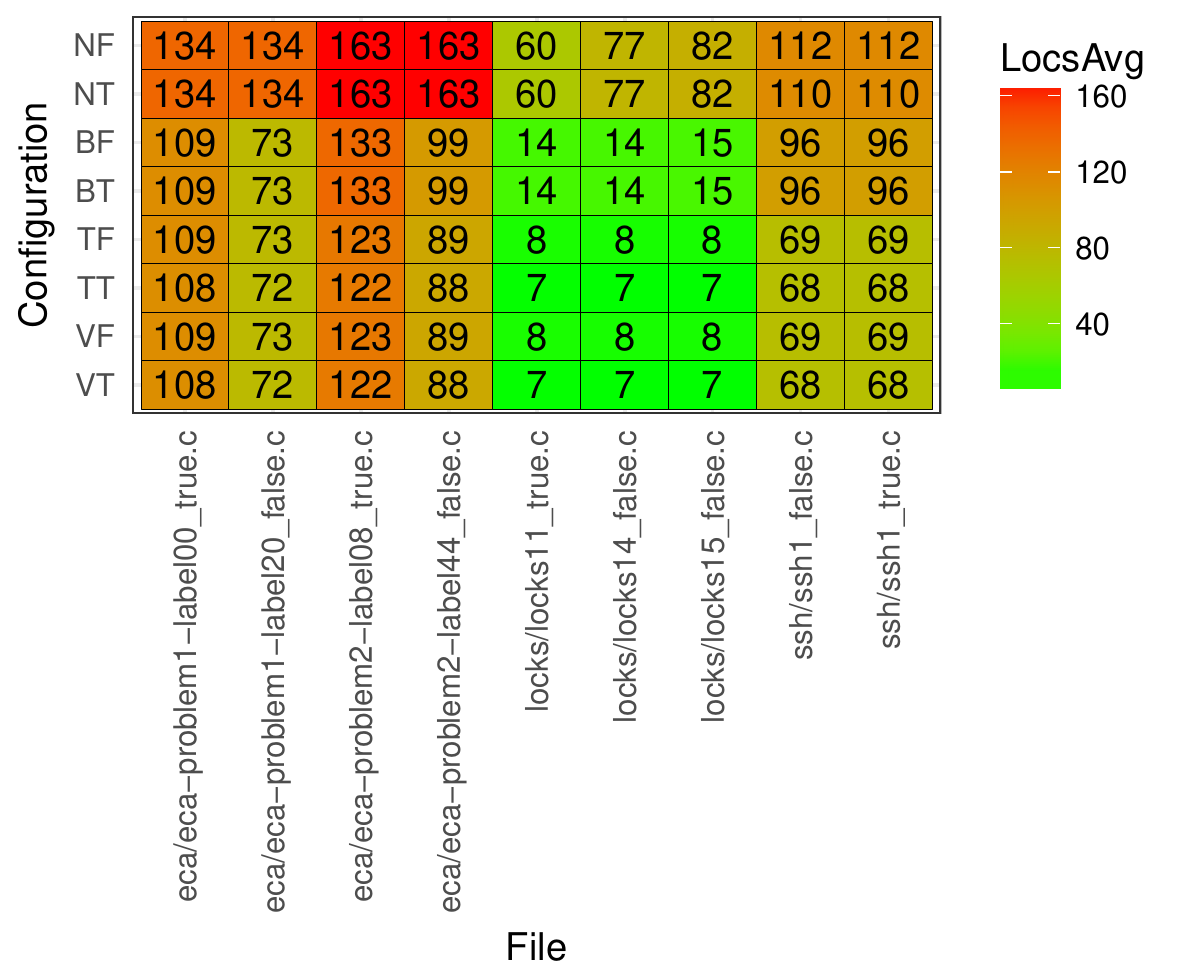}
	\caption{Heat map of average initial slice sizes.}
	\label{fig:slicer-cfa-size-heat}
\end{figure}

The heat map of the average initial slice sizes is shown in Figure~\ref{fig:slicer-cfa-size-heat}. The vertical axis lists the possible configurations without the search strategy value.
For example, \varvalue{NF} stands for slicing \varvalue{NONE}, optimizations \varvalue{FALSE}. 
Our calculations showed that the standard deviation of the slice sizes is $5.13$ at most, therefore displaying and evaluating averages has no significant distorting effect.
The first row shows CFA sizes without any slicing and optimization algorithms. Backward slicing reduces the CFA size considerably, even more so in the programs of the \textbf{locks} category.
Value and thin slicing allow even further reductions, except for the \textbf{eca} category. Since the assertions in the current programs do not contain any variables, the size of the initial CFAs are the same for both slicers.
We can also conclude that compiler optimizations have little to no effect on CFA size.
Nevertheless, the simplified expressions built by constant propagation may still allow some decrease in verification execution time.

\subsubsection{Impact of Slicing and Optimization on Verification Performance}

\begin{figure}[!htb]
	\centering
	\begin{subfigure}[b]{0.7\linewidth}		
		\centering
		\includegraphics[width=\linewidth]{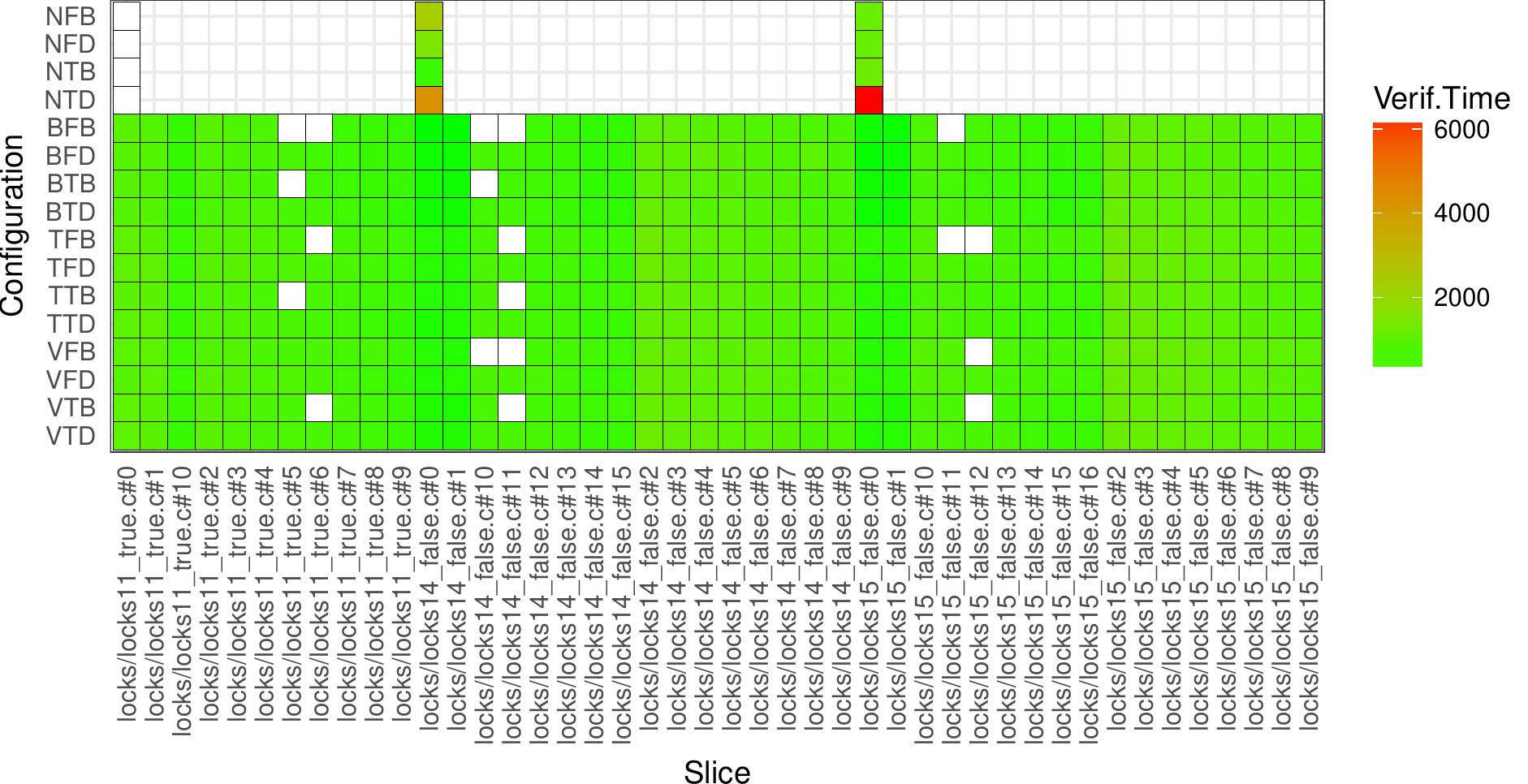}
		\caption{Verification time on the \textbf{locks} set.}
		\label{fig:slicer-cfa-ver-heat:a}
	\end{subfigure}
	\begin{subfigure}[b]{0.29\linewidth}
		\centering
		\includegraphics[width=\linewidth]{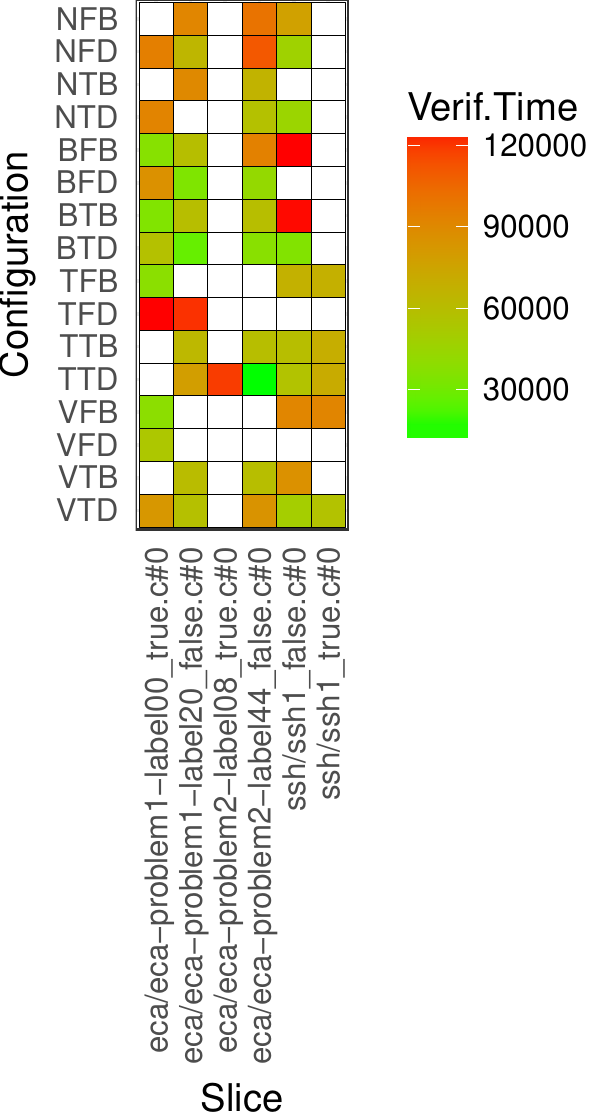}
		\caption{Verification time on the \textbf{eca} and \textbf{ssh} sets.}
		\label{fig:slicer-cfa-ver-heat:b}
	\end{subfigure}
	\caption{Heat map of verification time for slices with different configurations.}
	\label{fig:slicer-cfa-ver-heat}
\end{figure}

Figure~\ref{fig:slicer-cfa-ver-heat} shows a heat map of the verification execution time with different input configurations. White grids represent unsuccessful executions (due to timeouts).
The unfilled grids with light gray borders represent cases where a particular measurement data does not exist, because its corresponding slice has not been produced by the given slicer setting.
Figure~\ref{fig:slicer-cfa-ver-heat:a} depicts results on the \textbf{locks} task set. As it can be seen, the slices in this category mostly produce fairly similar execution times with different slicers.
Without slicing however, the verification execution time is quite large: for example, the \texttt{locks11\_true.c} ran into a timeout with all configurations that did not use slicing. For sliced programs, the verification time is rather promising. Timeouts on some slices with the BFS search strategy suggests that this search method can be inferior to the DFS strategy in a few cases.

The measurements for the \textbf{ssh} and \textbf{eca} categories (presented in~Figure~\ref{fig:slicer-cfa-ver-heat:b}) show really diverse results.
In most cases, slicing and optimizations decreased verifier execution time and allowed the verification of programs which have ran into timeout previously.
However, actual times vary through different slicing methods and search strategies. For example, the \texttt{eca-problem1-label00\_true.c} task performed better with the BFS search strategy in most cases, while
\texttt{eca-problem1-label20\_false.c} performed better with DFS. This supports the need for a configurable verification framework -- like the one presented in this paper, that is able to support these different slicing methods and search strategies. It is also interesting that there was a task (\texttt{eca-problem2-label08\_true.c}) that could only be verified using a single configuration (\varvalue{TTD}, i.e., thin slicing with optimizations, DFS search). This case could be the subject of a more detailed analysis in the future.

\begin{figure}[!htb]
	\centering
	\begin{subfigure}[b]{0.49\linewidth}		
		\includegraphics[width=\linewidth]{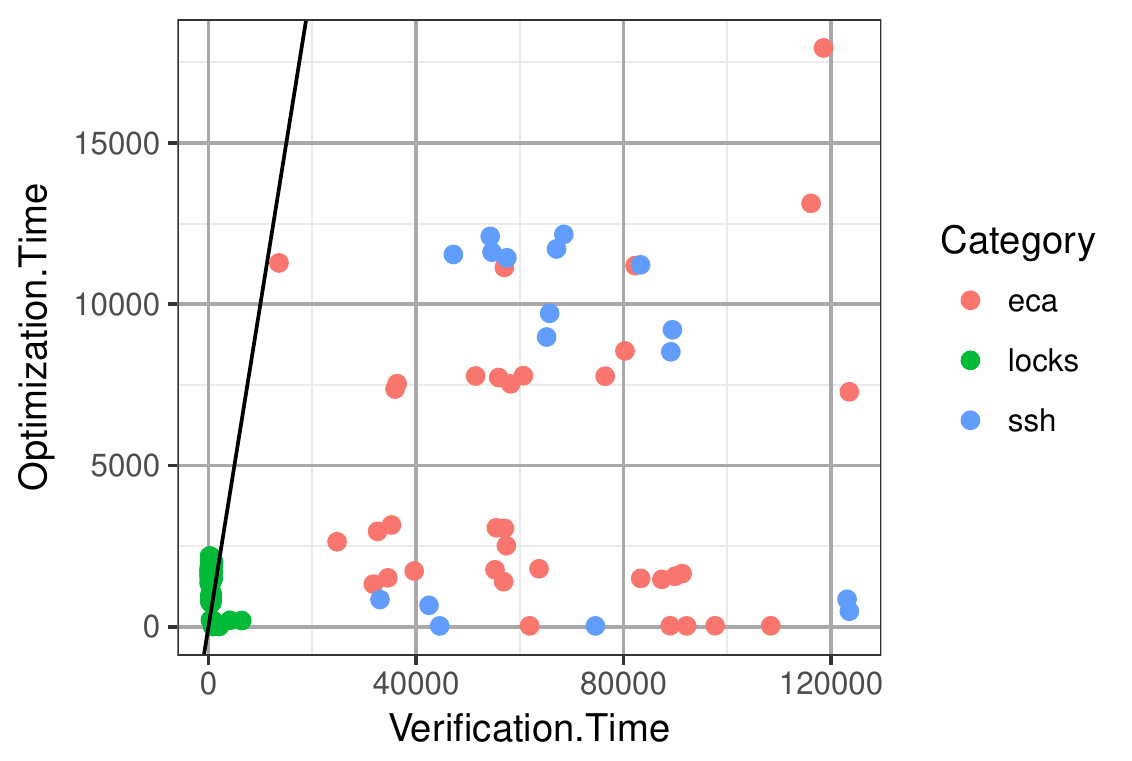}
		\caption{Verification and optimization time by categories.}
		\label{fig:opt-ver-time-category}
	\end{subfigure}
	\begin{subfigure}[b]{0.49\linewidth}	
		\includegraphics[width=\linewidth]{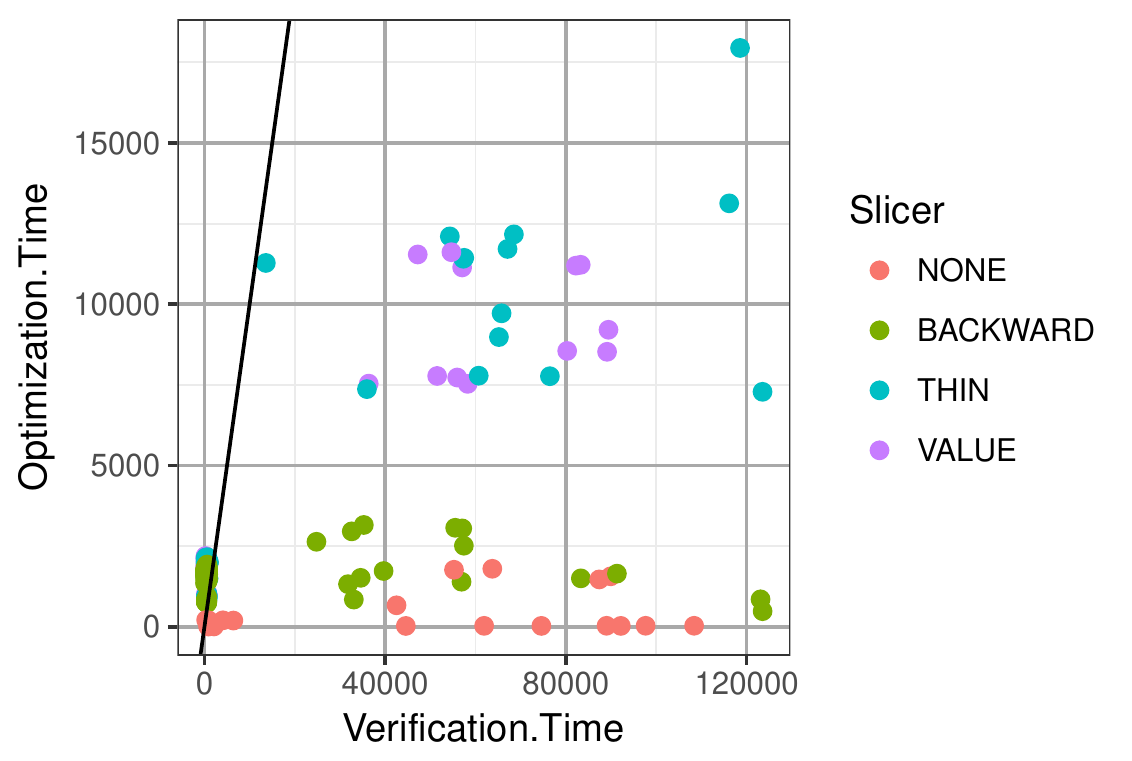}
		\caption{Verification and optimization time by slicers.}
		\label{fig:opt-ver-time-slicer}
	\end{subfigure}	
	\caption{Comparison of verification time and optimization time.}
	\label{fig:opt-ver-slice-time}
\end{figure}

We also compared verification time and optimization time. Their comparison, grouped by program categories is shown in Figure~\ref{fig:opt-ver-time-category}.
As it can be seen, optimization takes roughly as much time as verification on small programs (from \textbf{locks}).
For larger programs however, optimization time is negligible compared to the verifier's execution time.
Note that optimization time also includes all time spent on slicing and slice refinement.
Figure~\ref{fig:opt-ver-time-slicer} shows the same comparison grouped by slicer types.
We can conclude that without slicing or with backward slicing, the optimization time is rather fast, especially when compared to verification time.
For value slicing, the optimization time increases significantly due to the refinements. For thin slicing this time increases slightly further, except for a few exceptions,
where thin slicing produced much greater optimization times.

\paragraph{Threats to validity.} External validity of our results is currently limited to our tool and the selected categories of SV-COMP. Selecting more models and tools from different sources could improve external validity. Internal validity was increased by a fully automated measurement environment but it could also be improved by for example repeating the same measurements multiple times.
\section{Conclusions}
\label{sec:conclusions}

In this paper, we have presented and evaluated a framework for transforming C programs to control flow automata, enhanced by optimization transformations known from compiler design and different program slicing methods (namely backward, thin, and value slicing).
The resulting models are verified using the CEGAR model checking algorithm.
We also performed an experiment by evaluating the performance of the different slicing methods and verification configurations.
Our results show that the effectiveness of a certain slicing algorithm varies between certain input programs.
We also concluded that in some cases a certain slicing method performs better than the others.
All this information supports the need for a configurable framework -- like the one presented in this paper, which includes several slicing and optimization algorithms.

\paragraph{Future work.}
Our framework has several opportunities for improvement.
The range of supported features of the C language could be extended with for example arrays, pointers or structs.
Other slicing algorithms could be included in the workflow, such as interprocedural slicing~\cite{Horwitz90}. 
Adding support for the LLVM~IR~\cite{LLVM} would extend the range of supported languages and would also implicitly add multiple fine-tuned optimizations into the workflow.
Moreover, the effects of size reduction could be evaluated on a wider range of verifier configurations, e.g.\ using different abstract domains or refinement strategies.

\bibliographystyle{eptcs}
\bibliography{generic}

\begin{thebibliography}{10}
\providecommand{\bibitemdeclare}[2]{}
\providecommand{\surnamestart}{}
\providecommand{\surnameend}{}
\providecommand{\urlprefix}{Available at }
\providecommand{\url}[1]{\texttt{#1}}
\providecommand{\href}[2]{\texttt{#2}}
\providecommand{\urlalt}[2]{\href{#1}{#2}}
\providecommand{\doi}[1]{doi:\urlalt{http://dx.doi.org/#1}{#1}}
\providecommand{\bibinfo}[2]{#2}

\bibitemdeclare{book}{DragonBook}
\bibitem{DragonBook}
\bibinfo{author}{Alfred~V. \surnamestart Aho\surnameend}, \bibinfo{author}{Ravi
  \surnamestart Sethi\surnameend} \& \bibinfo{author}{Jeffrey~D. \surnamestart
  Ullman\surnameend} (\bibinfo{year}{1986}): \emph{\bibinfo{title}{Compilers:
  Principles, Techniques, and Tools}}.
\newblock \bibinfo{publisher}{Addison-Wesley Longman Publishing Co., Inc.},
  \bibinfo{address}{Boston, MA, USA}.

\bibitemdeclare{article}{SLAM}
\bibitem{SLAM}
\bibinfo{author}{Thomas \surnamestart Ball\surnameend} \&
  \bibinfo{author}{Sriram~K. \surnamestart Rajamani\surnameend}
  (\bibinfo{year}{2002}): \emph{\bibinfo{title}{The {SLAM} Project: Debugging
  System Software via Static Analysis}}.
\newblock {\sl \bibinfo{journal}{SIGPLAN Not.}}
  \bibinfo{volume}{37}(\bibinfo{number}{1}), pp. \bibinfo{pages}{1--3},
  \doi{10.1145/565816.503274}.

\bibitemdeclare{incollection}{Beyer16}
\bibitem{Beyer16}
\bibinfo{author}{Dirk \surnamestart Beyer\surnameend} (\bibinfo{year}{2016}):
  \emph{\bibinfo{title}{Reliable and Reproducible Competition Results with
  BenchExec and Witnesses (Report on {SV-COMP} 2016)}}.
\newblock In: {\sl \bibinfo{booktitle}{Tools and Algorithms for the
  Construction and Analysis of Systems}}, {\sl \bibinfo{series}{Lecture Notes
  in Computer Science}} \bibinfo{volume}{9636}, \bibinfo{publisher}{Springer},
  pp. \bibinfo{pages}{887--904}, \doi{10.1007/978-3-662-49674-9_55}.

\bibitemdeclare{inproceedings}{Beyer09}
\bibitem{Beyer09}
\bibinfo{author}{Dirk \surnamestart Beyer\surnameend},
  \bibinfo{author}{Alessandro \surnamestart Cimatti\surnameend},
  \bibinfo{author}{Alberto \surnamestart Griggio\surnameend},
  \bibinfo{author}{M.~Erkan \surnamestart Keremoglu\surnameend} \&
  \bibinfo{author}{Roberto \surnamestart Sebastiani\surnameend}
  (\bibinfo{year}{2009}): \emph{\bibinfo{title}{Software model checking via
  large-block encoding}}.
\newblock In: {\sl \bibinfo{booktitle}{Proceedings of the 2009 Conference on
  Formal Methods in Computer-Aided Design}}, \bibinfo{publisher}{IEEE}, pp.
  \bibinfo{pages}{25--32}, \doi{10.1109/FMCAD.2009.5351147}.

\bibitemdeclare{incollection}{Beyer16smt}
\bibitem{Beyer16smt}
\bibinfo{author}{Dirk \surnamestart Beyer\surnameend} \&
  \bibinfo{author}{Matthias \surnamestart Dangl\surnameend}
  (\bibinfo{year}{2016}): \emph{\bibinfo{title}{{SMT}-based Software Model
  Checking: An Experimental Comparison of Four Algorithms}}.
\newblock In: {\sl \bibinfo{booktitle}{Verified Software. Theories, Tools, and
  Experiments}}, {\sl \bibinfo{series}{Lecture Notes in Computer Science}}
  \bibinfo{volume}{9971}, \bibinfo{publisher}{Springer}, pp.
  \bibinfo{pages}{181--198}, \doi{10.1007/978-3-319-48869-1_14}.

\bibitemdeclare{article}{Blast}
\bibitem{Blast}
\bibinfo{author}{Dirk \surnamestart Beyer\surnameend},
  \bibinfo{author}{Thomas~A. \surnamestart Henzinger\surnameend},
  \bibinfo{author}{Ranjit \surnamestart Jhala\surnameend} \&
  \bibinfo{author}{Rupak \surnamestart Majumdar\surnameend}
  (\bibinfo{year}{2007}): \emph{\bibinfo{title}{The software model checker
  {B}last}}.
\newblock {\sl \bibinfo{journal}{International Journal on Software Tools for
  Technology Transfer}} \bibinfo{volume}{9}(\bibinfo{number}{5}), pp.
  \bibinfo{pages}{505--525}, \doi{10.1007/s10009-007-0044-z}.

\bibitemdeclare{incollection}{CPAChecker}
\bibitem{CPAChecker}
\bibinfo{author}{Dirk \surnamestart Beyer\surnameend},
  \bibinfo{author}{Thomas~A. \surnamestart Henzinger\surnameend} \&
  \bibinfo{author}{Gr{\'e}gory \surnamestart Th{\'e}oduloz\surnameend}
  (\bibinfo{year}{2007}): \emph{\bibinfo{title}{Configurable Software
  Verification: Concretizing the Convergence of Model Checking and Program
  Analysis}}.
\newblock In: {\sl \bibinfo{booktitle}{Computer Aided Verification}}, {\sl
  \bibinfo{series}{Lecture Notes in Computer Science}} \bibinfo{volume}{4590},
  \bibinfo{publisher}{Springer}, pp. \bibinfo{pages}{504--518},
  \doi{10.1007/978-3-540-73368-3_51}.

\bibitemdeclare{incollection}{Beyer15refinement}
\bibitem{Beyer15refinement}
\bibinfo{author}{Dirk \surnamestart Beyer\surnameend}, \bibinfo{author}{Stefan
  \surnamestart L{\"o}we\surnameend} \& \bibinfo{author}{Philipp \surnamestart
  Wendler\surnameend} (\bibinfo{year}{2015}): \emph{\bibinfo{title}{Refinement
  Selection}}.
\newblock In: {\sl \bibinfo{booktitle}{Model Checking Software: 22nd
  International Symposium, SPIN 2015, Stellenbosch, South Africa, August 24-26,
  2015, Proceedings}}, {\sl \bibinfo{series}{Lecture Notes in Computer
  Science}} \bibinfo{volume}{9232}, \bibinfo{publisher}{Springer}, pp.
  \bibinfo{pages}{20--38}, \doi{10.1007/978-3-319-23404-5_3}.

\bibitemdeclare{article}{Binkley07}
\bibitem{Binkley07}
\bibinfo{author}{David \surnamestart Binkley\surnameend},
  \bibinfo{author}{Nicolas \surnamestart Gold\surnameend} \&
  \bibinfo{author}{Mark \surnamestart Harman\surnameend}
  (\bibinfo{year}{2007}): \emph{\bibinfo{title}{An Empirical Study of Static
  Program Slice Size}}.
\newblock {\sl \bibinfo{journal}{ACM Transactions on Software Engineering and
  Methodology}} \bibinfo{volume}{16}(\bibinfo{number}{2}),
  \doi{10.1145/1217295.1217297}.

\bibitemdeclare{article}{SLAB}
\bibitem{SLAB}
\bibinfo{author}{Ingo \surnamestart Br\"{u}ckner\surnameend},
  \bibinfo{author}{Klaus \surnamestart Dr\"{a}ger\surnameend},
  \bibinfo{author}{Bernd \surnamestart Finkbeiner\surnameend} \&
  \bibinfo{author}{Heike \surnamestart Wehrheim\surnameend}
  (\bibinfo{year}{2008}): \emph{\bibinfo{title}{Slicing Abstractions}}.
\newblock {\sl \bibinfo{journal}{Fundamenta Informaticae}}
  \bibinfo{volume}{89}(\bibinfo{number}{4}), pp. \bibinfo{pages}{369--392}.

\bibitemdeclare{article}{clarke03}
\bibitem{clarke03}
\bibinfo{author}{Edmund \surnamestart Clarke\surnameend}, \bibinfo{author}{Orna
  \surnamestart Grumberg\surnameend}, \bibinfo{author}{Somesh \surnamestart
  Jha\surnameend}, \bibinfo{author}{Yuan \surnamestart Lu\surnameend} \&
  \bibinfo{author}{Helmut \surnamestart Veith\surnameend}
  (\bibinfo{year}{2003}): \emph{\bibinfo{title}{Counterexample-guided
  abstraction refinement for symbolic model checking}}.
\newblock {\sl \bibinfo{journal}{Journal of the ACM}}
  \bibinfo{volume}{50}(\bibinfo{number}{5}), pp. \bibinfo{pages}{752--794},
  \doi{10.1145/876638.876643}.

\bibitemdeclare{article}{clarke04}
\bibitem{clarke04}
\bibinfo{author}{Edmund \surnamestart Clarke\surnameend},
  \bibinfo{author}{Anubhav \surnamestart Gupta\surnameend} \&
  \bibinfo{author}{Ofer \surnamestart Strichman\surnameend}
  (\bibinfo{year}{2004}): \emph{\bibinfo{title}{{SAT}-based
  counterexample-guided abstraction refinement}}.
\newblock {\sl \bibinfo{journal}{IEEE Trans. on Computer-Aided Design of
  Integrated Circuits and Systems}} \bibinfo{volume}{23}(\bibinfo{number}{7}),
  pp. \bibinfo{pages}{1113--1123}, \doi{10.1109/TCAD.2004.829807}.

\bibitemdeclare{incollection}{SATABS}
\bibitem{SATABS}
\bibinfo{author}{Edmund \surnamestart Clarke\surnameend},
  \bibinfo{author}{Daniel \surnamestart Kroening\surnameend},
  \bibinfo{author}{Natasha \surnamestart Sharygina\surnameend} \&
  \bibinfo{author}{Karen \surnamestart Yorav\surnameend}
  (\bibinfo{year}{2005}): \emph{\bibinfo{title}{{SATABS}: {SAT}-Based Predicate
  Abstraction for {ANSI-C}}}.
\newblock In: {\sl \bibinfo{booktitle}{Tools and Algorithms for the
  Construction and Analysis of Systems}}, {\sl \bibinfo{series}{Lecture Notes
  in Computer Science}} \bibinfo{volume}{3440}, \bibinfo{publisher}{Springer},
  pp. \bibinfo{pages}{570--574}, \doi{10.1007/978-3-540-31980-1_40}.

\bibitemdeclare{article}{Ferrante87}
\bibitem{Ferrante87}
\bibinfo{author}{Jeanne \surnamestart Ferrante\surnameend},
  \bibinfo{author}{Karl~J. \surnamestart Ottenstein\surnameend} \&
  \bibinfo{author}{Joe~D. \surnamestart Warren\surnameend}
  (\bibinfo{year}{1987}): \emph{\bibinfo{title}{The Program Dependence Graph
  and Its Use in Optimization}}.
\newblock {\sl \bibinfo{journal}{ACM Trans. Program. Lang. Syst.}}
  \bibinfo{volume}{9}(\bibinfo{number}{3}), pp. \bibinfo{pages}{319--349},
  \doi{10.1145/24039.24041}.

\bibitemdeclare{incollection}{graf97}
\bibitem{graf97}
\bibinfo{author}{Susanne \surnamestart Graf\surnameend} \&
  \bibinfo{author}{Hassen \surnamestart Saidi\surnameend}
  (\bibinfo{year}{1997}): \emph{\bibinfo{title}{Construction of abstract state
  graphs with {PVS}}}.
\newblock In: {\sl \bibinfo{booktitle}{Computer Aided Verification}}, {\sl
  \bibinfo{series}{Lecture Notes in Computer Science}} \bibinfo{volume}{1254},
  \bibinfo{publisher}{Springer}, pp. \bibinfo{pages}{72--83},
  \doi{10.1007/3-540-63166-6_10}.

\bibitemdeclare{incollection}{forte2016}
\bibitem{forte2016}
\bibinfo{author}{\'Akos \surnamestart Hajdu\surnameend},
  \bibinfo{author}{Tam\'as \surnamestart T\'oth\surnameend},
  \bibinfo{author}{Andr\'as \surnamestart V\"or\"os\surnameend} \&
  \bibinfo{author}{Istv\'an \surnamestart Majzik\surnameend}
  (\bibinfo{year}{2016}): \emph{\bibinfo{title}{A Configurable {CEGAR}
  Framework with Interpolation-based Refinements}}.
\newblock In: {\sl \bibinfo{booktitle}{Formal Techniques for Distributed
  Objects, Components and Systems}}, {\sl \bibinfo{series}{Lecture Notes in
  Computer Science}} \bibinfo{volume}{9688}, \bibinfo{publisher}{Springer}, pp.
  \bibinfo{pages}{158--174}, \doi{10.1007/978-3-319-39570-8_11}.

\bibitemdeclare{article}{Horwitz90}
\bibitem{Horwitz90}
\bibinfo{author}{Susan \surnamestart Horwitz\surnameend},
  \bibinfo{author}{Thomas \surnamestart Reps\surnameend} \&
  \bibinfo{author}{David \surnamestart Binkley\surnameend}
  (\bibinfo{year}{1990}): \emph{\bibinfo{title}{Interprocedural Slicing Using
  Dependence Graphs}}.
\newblock {\sl \bibinfo{journal}{ACM Transactions on Programming Languages and
  Systems}} \bibinfo{volume}{12}(\bibinfo{number}{1}), pp.
  \bibinfo{pages}{26--60}, \doi{10.1145/77606.77608}.

\bibitemdeclare{incollection}{Kumar15}
\bibitem{Kumar15}
\bibinfo{author}{Shrawan \surnamestart Kumar\surnameend},
  \bibinfo{author}{Amitabha \surnamestart Sanyal\surnameend} \&
  \bibinfo{author}{Uday~P. \surnamestart Khedker\surnameend}
  (\bibinfo{year}{2015}): \emph{\bibinfo{title}{Value Slice: A New Slicing
  Concept for Scalable Property Checking}}.
\newblock In: {\sl \bibinfo{booktitle}{Tools and Algorithms for the
  Construction and Analysis of Systems}}, {\sl \bibinfo{series}{Lecture Notes
  in Computer Science}} \bibinfo{volume}{9035}, \bibinfo{publisher}{Springer},
  pp. \bibinfo{pages}{101--115}, \doi{10.1007/978-3-662-46681-0_7}.

\bibitemdeclare{inproceedings}{LLVM}
\bibitem{LLVM}
\bibinfo{author}{Chris \surnamestart Lattner\surnameend} \&
  \bibinfo{author}{Vikram \surnamestart Adve\surnameend}
  (\bibinfo{year}{2004}): \emph{\bibinfo{title}{{LLVM}: A Compilation Framework
  for Lifelong Program Analysis \& Transformation}}.
\newblock In: {\sl \bibinfo{booktitle}{Proceedings of the 2004 International
  Symposium on Code Generation and Optimization}}, \bibinfo{publisher}{IEEE},
  pp. \bibinfo{pages}{75--86}.

\bibitemdeclare{inproceedings}{leucker15}
\bibitem{leucker15}
\bibinfo{author}{Martin \surnamestart Leucker\surnameend},
  \bibinfo{author}{Grigory \surnamestart Markin\surnameend} \&
  \bibinfo{author}{Martin~R. \surnamestart Neuh\"au{\ss}er\surnameend}
  (\bibinfo{year}{2015}): \emph{\bibinfo{title}{A New Refinement Strategy for
  {CEGAR}-Based Industrial Model Checking}}.
\newblock In: {\sl \bibinfo{booktitle}{Hardware and Software: Verification and
  Testing}}, {\sl \bibinfo{series}{Lecture Notes in Computer Science}}
  \bibinfo{volume}{9434}, \bibinfo{publisher}{Springer}, pp.
  \bibinfo{pages}{155--170}, \doi{10.1007/978-3-319-26287-1_10}.

\bibitemdeclare{incollection}{mcmillan05}
\bibitem{mcmillan05}
\bibinfo{author}{Kenneth~L. \surnamestart McMillan\surnameend}
  (\bibinfo{year}{2005}): \emph{\bibinfo{title}{Applications of {C}raig
  Interpolants in Model Checking}}.
\newblock In: {\sl \bibinfo{booktitle}{Tools and Algorithms for the
  Construction and Analysis of Systems}}, {\sl \bibinfo{series}{Lecture Notes
  in Computer Science}} \bibinfo{volume}{3440}, \bibinfo{publisher}{Springer},
  pp. \bibinfo{pages}{1--12}, \doi{10.1007/978-3-540-31980-1_1}.

\bibitemdeclare{inproceedings}{Sridharan07}
\bibitem{Sridharan07}
\bibinfo{author}{Manu \surnamestart Sridharan\surnameend},
  \bibinfo{author}{Stephen~J. \surnamestart Fink\surnameend} \&
  \bibinfo{author}{Rastislav \surnamestart Bodik\surnameend}
  (\bibinfo{year}{2007}): \emph{\bibinfo{title}{Thin Slicing}}.
\newblock In: {\sl \bibinfo{booktitle}{Proceedings of the 28th ACM SIGPLAN
  Conference on Programming Language Design and Implementation}},
  \bibinfo{publisher}{ACM}, pp. \bibinfo{pages}{112--122}.

\bibitemdeclare{inproceedings}{vizel09}
\bibitem{vizel09}
\bibinfo{author}{Yakir \surnamestart Vizel\surnameend} \& \bibinfo{author}{Orna
  \surnamestart Grumberg\surnameend} (\bibinfo{year}{2009}):
  \emph{\bibinfo{title}{Interpolation-sequence based model checking}}.
\newblock In: {\sl \bibinfo{booktitle}{Proceedings of the 2009 Conference on
  Formal Methods in Computer-Aided Design}}, \bibinfo{organization}{IEEE}, pp.
  \bibinfo{pages}{1--8}, \doi{10.1109/FMCAD.2009.5351148}.

\bibitemdeclare{inproceedings}{Weiser81}
\bibitem{Weiser81}
\bibinfo{author}{Mark \surnamestart Weiser\surnameend} (\bibinfo{year}{1981}):
  \emph{\bibinfo{title}{Program Slicing}}.
\newblock In: {\sl \bibinfo{booktitle}{Proceedings of the 5th International
  Conference on Software Engineering}}, \bibinfo{publisher}{IEEE}, pp.
  \bibinfo{pages}{439--449}.

\end{thebibliography}


\end{document}